\documentstyle[12pt,psfig]{article}

\everyjob{\typeout{MuTeX.sty: LaTeX Modification by EPM.}}
\immediate\write10{MuTeX.sty: LaTeX Modification by EPM.}

\makeatletter   

\input{ifthen.sty}

\newtheorem {theorem} {Theorem} [section]

\newtheorem {lemma} [theorem] {Lemma}

\newtheorem {Canham threshold} [theorem] {Canham Threshold}

\def\theoremstyle#1#2{\def\@@theoremheadstyle{#1}
                      \def\@@theorembodystyle{#2}}
\def\@@theoremheadstyle{\sc}
\def\@@theorembodystyle{\rm}
\def\@begintheorem#1#2{\@@theorembodystyle 
                       \trivlist 
		       \item[\hskip 
                             \labelsep{\@@theoremheadstyle #1\ #2}]}
\def\@opargbegintheorem#1#2#3{\@@theorembodystyle 
                              \trivlist 
			       \item[\hskip 
				  \labelsep{\@@theoremheadstyle #1\ #2\ (#3)}]}

 \def\@@pc{\bf}

 \newcommand {\pcodestyle}[1] {\def\@@pc{#1}}  

 {
 \def\PROGRAM		{{\@@pc program\ }}
 
 \def\PROCEDURE		{{\@@pc procedure\ }}
 \def\FUNCTION		{{\@@pc function\ }}
 \def\LOCAL		{{\@@pc local\ }}
 \def\GLOBAL		{{\@@pc global\ }}
 \def\RETURNS		{{\@@pc returns\ }}
 \def\RETURN		{{\@@pc return\ }}
 \def\BEGIN		{{\@@pc begin\ }}
 \def\END		{{\@@pc end\ }}
 \def\IF			{{\@@pc if\ }}
 \def\THEN		{{\@@pc then\ }}
 \def\ELSE		{{\@@pc else\ }}
 \def\REPEAT		{{\@@pc repeat\ }}
 \def\UNTIL		{{\@@pc until\ }}
 \def\WHILE		{{\@@pc while\ }}
 \def\DO			{{\@@pc do\ }}
 \def\FOR		{{\@@pc for\ }}
 \def\TO			{{\@@pc to\ }}
 \def\DOWN		{{\@@pc down\ }}
 \def\NEXT		{{\@@pc next\ }}
 \begin{center}
  \begin{minipage}{.9\textwidth}
   \small
    \begin{tabbing}
 }%
 {  \end{tabbing}%
   \end{minipage}%
  \end{center}%
 }

			   {\end{quotation}}

			 {  \end{minipage}%
			   \end{center}%
			  \end{description}%
			 }

				{ \end{minipage}
				 \end{center}%
				}

\def\thebibliography#1{\section*{References}\list
 {[\arabic{enumi}]}{\settowidth\labelwidth{[#1]}\leftmargin\labelwidth
 \advance\leftmargin\labelsep
 \usecounter{enumi}}
 \def\newblock{\hskip .11em plus .33em minus -.07em}
 \sloppy
 \sfcode`\.=1000\relax}

\newsavebox{\ProofSym}
\savebox{\ProofSym}{%
  \begin{picture}(10,10)
    \put(0,0){\framebox(9,9){}}
    \put(0,3){\framebox(6,6){}}
  \end{picture}}
\newcommand{\eop}{\hfill\usebox{\ProofSym}}
\newenvironment{proof}{\noindent {\sc Proof.\/}}{\eop\par\vspace{0.3cm}}

\theoremstyle{\sc}{\rm}

\begin{document}

\bibliographystyle{plain}


\begin{center}
      {\Large\bf
       ~ \\ 
       ~ \\
       ~ \\
       ~ \\
On multiple connectedness of regions visible due to multiple diffuse
reflections
}
      ~\\
      ~\\
      ~\\
        
	{\large Sudebkumar Prasant Pal\footnote{A part of this work was done
when this author was visiting the University of Miami, Coral Gables, Florida,
USA.}}\\
		Department of Computer Science and Engineering,\\
		Indian Institute of Technology, Kharagpur\\
		Kharagpur 721302, India.\\
        email: spp@cse.iitkgp.ernet.in\\ 
          url:- http://www.angelfire.com/or/sudebkumar\\  
        {\large Dilip Sarkar}\\
        Department of Computer Science\\
        University of Miami, Coral Gables\\
        Miami, FL 33124, USA\\
        email: sarkar@cs.miami.edu\\
      ~\\
      ~\\
\end{center}

\begin{abstract}

It is known that the region $V(s)$
of a simple polygon $P$, directly visible (illuminable) from 
an internal point $s$,
is simply connected. Aronov et al. \cite{addpp981} 
established that the region $V_1(s)$
of a simple polygon visible from an internal point $s$ due 
to at most one diffuse reflection on the boundary of 
the polygon $P$, is also simply 
connected. In this paper we establish that the region 
$V_2(s)$, visible from $s$ due  
to at most two diffuse reflections may 
be multiply connected; we demonstrate the construction of an $n$-sided
simple polygon with a point $s$ inside it so that and the 
region of $P$ visible
from $s$ after at most two diffuse 
reflections is multiple connected.

\end{abstract}



\section{Introduction}

In recent years, visibility problems have been 
studied extensively (see
\cite{o87,s92}). Visibility computations abound in computer graphics, 
motion planning,
robotics and computer vision.
Two points inside a simple 
polygon are mutually $visible$ if the line segment joining them 
is not obstructed by any edge of the polygon.
Several algorithms exist for computing 
the region visible from a point light 
source inside a simple 
polygon \cite{ea81,l83,ghlst87}. 
The problems of computing 
the region of a simple polygon which is
{\it weakly visible} from an internal segment 
\cite{ll86b,ghlst87},  or a convex set \cite{g91},
are also well studied. A point $p$ inside a simple polygon $P$ is said to be
{\it weakly visible} from an edge $e$ of the polygon if $p$ is visible from
some point in the {\it interior} of $e$. 
Certain portions of the polygon that are 
not directly illuminated from the source may become visible due
to one or more reflections on the bounding edges of the polygon. 
Reflection at a point is called {\it specular} if the
reflected ray is directed as per the standard law of reflection;
the angle of incidence is the same as the angle of reflection.
Most reflecting surfaces cause another type of reflection, called 
{\it diffuse} reflection, where
light incident at a point is reflected in all 
possible interior directions. 
We assume that light incident at 
a vertex is absorbed and not reflected.

Throughout this paper we consider the region visible from a 
point source $s$ due to multiple reflections inside a simple polygon $P$.
Such visible regions are also called {\it visibility polygons}.

\subsection{Preliminaries and notation}
\label{def} 

Let $P$
be a simple polygon with no three collinear
vertices. Let $int(X)$ and $bd(X)$ denote the relative interior and
boundary of a region $X$ of $P$, respectively. If $p$ and $q$ are two points on
$bd(P)$, let $bd(p,q)$  denote the part of $bd(P)$
traversed from $p$ to $q$ in counterclockwise order, 
keeping $int(P)$ to the left 
of the irection of traversal.
Two points in $P$ are {\it (mutually) visible} if the
interior of the line segment joining
them lies in $int(P)$. 
A point $y$ is {\it visible  under diffuse reflection}
from a point $x$, if there exists a point $p$ lying in the interior
of an edge of $P$, such that
$p$  is visible from both $x$ and $y$.

\begin{figure}[htbp]
\centerline{\psfig{figure=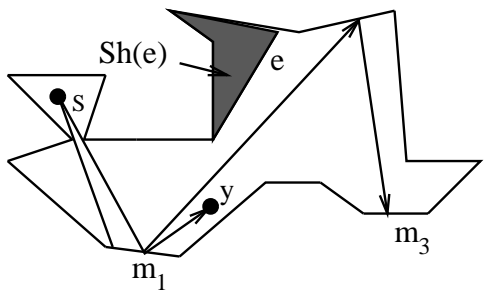}}
\caption{Visibility with multiple diffuse reflections.}
\label{kpath}
\end{figure}

Let $S$ denote a point light source inside $P$.
A point $y$ is $1$-visible from $S$ if $y$ and $S$ are mutually visible.
For $k>1$, a point $y(=p_k)$ is {\it $k$-visible} 
from a point $S(=p_0)$, if there exists  points
$p_1,p_2,...,p_{k-1}$ in the interiors of edges of $P$,
such that $p_i$ and $p_{i+1}$ are mutually visible for $0\leq i\leq k-1$.
We say that the ray emitted from $S$
reaches $y$ after $k-1$ {\it diffuse reflections} 
at points $p_1,p_2,...,p_{k-1}$.
In Figure \ref{kpath}, $y$ is {\em 2-visible} from $S$;
$y$ is visible from $S$ after one diffuse reflection. 
We assume that the light incident at a vertex is absorbed and not reflected
further.

Let $V(s)$ denote the portion of $P$ visible from $s$.
We know that $V(s)$ is simply connected, has $O(n)$ edges, 
and at most one edge of $V(s)$ lies on each edge of $P$ \cite{ea81,ghlst87}.
For a point $S\in P$, let $V(S)=V(P,S)$ denote the polygonal region
consisting of points in $P$ that are directly visible from $S$. 
For $k \geq 0$, let $V_k(S)$ denote the polygonal region consisting 
of points that are $l$-visible from $S$, for some $1\leq l \leq k+1$. 
$V_k(s)$ is the set of points that receive light from $s$ after
at most $k$ diffuse reflections. We have $V_0(s)=V(s)$.

\subsection{Combinatorial complexity and multiple connectedness 
of regions visible due to diffuse reflections}

Regions visible due to reflections inside simple polygons
were first studied by Aronov {\em et al.} \cite{addpp981}; 
a tight $\Theta(n^2)$ bound was established on the number of edges 
of visibility polygons when at most
one diffuse reflection is permitted. The authors showed that such visibility
polygons were simply connected. In sharp contrast, 
regions visible due to at most one specular reflection were shown to be
multiply connected, with combinatorial complexity 
$\Theta(n^2)$ 
\cite{addpp981}. 
(The combinatorial complexity of a visibility polygon is the number of edges
defining its boundary.) 
Indeed, the number of {\it holes} 
(also called {\it blind spots})
in such a 
visibility polygon (with at most one specular reflection) is 
$\Theta(n^2)$  \cite{addpp981}. 

The case
where at most $k$ specular reflections are permitted, is dealt with in  
a subsequent paper \cite{addpp982}; 
an $O(n^{2k})$ upper bound and a lower bound of
$\Omega((n/k-\Theta (1))^{2k})$ 
on the combinatorial complexity of the visibility polygons was established
when at most a constant number 
$k$ of specular reflections are permitted. The number of holes too 
obeys these bounds.
The upper bound on the combinatorial complexity of visibility polygons
improves to 
$O(n^{2\lceil (k+1)/2\rceil +1})$,  
when specular reflections are replaced by diffuse reflections \cite{ppd98}. 
Note that the improvement is by an asymptotic square root factor for large $k$.
This is not surprizing because diffuse reflection spreads light much more 
widely than restricted pencils do in 
the case of  specular reflections. 

The
best lower bound on the combinatorial complexity for visibility 
polygons $V_k(s)$ (with at most $k$ diffuse reflections),
for any $k\geq 1$, remains just $\Omega(n^2)$ to date (see 
\cite{addpp981,ppd98}). 
One interesting and important problem concerning multiple diffuse
reflections is that of bridging this gap between the very high upper bound of 
\cite{ppd98} and
the $\Omega (n^2)$ lower bound of \cite{addpp981,ppd98}.
One way to approach this open problem is to study the structure of $V_k(s)$ and
bound the number of holes in it. Showing that the number of holes is not as 
high as in the case of spcular reflections (see \cite{addpp982}), may help  
improving the upper bound. 
The presence of holes in $V_2(s)$ (and in $V_k(s)$ for $k\geq 1$, in general) 
was not known until we demonstrated
the existence of holes in $V_k(s)$, $k\geq 1$, in 
\cite{sppdstrmay2001}. This paper presents 
essentially the construction in \cite{sppdstrmay2001} of an $n$-sided simple 
polygon $P$ with a point $s$ located inside $P$ so that $V_2(s)$ has a hole.
We construct a simple 
polygon $P$ and place a point $s$ inside it so that the portion of the 
polygon $P$ visible from $s$ due to at most two diffuse reflections 
does not include the interior
of a triangle $tqr$ (see Figures \ref{figex3} and \ref{figex2}). 
Here, the triangle $tqr$ is totally contained inside $P$
and the boundary of triangle $tqr$ as well as its exterior 
neighbourhood are visible from $s$ after at most two diffuse reflections.

An interesting problem is to find asympototic bounds on the number of holes 
in $V_2(s)$. We believe that there can not be more than a linear number of 
holes. Such a linear bound on the number of holes in $V_k(s)$ for any 
$k\geq 1$ may improve the upper bound on the 
combinatorial complexity of $V_k(s)$ over that in \cite{ppd98}.

The existence of holes in $V_2(s)$ is a bit surprising. Diffuse reflection, 
as the term suggests, results in bouncing back of incident light 
in all possible directions from the point of reflection. Clearly, a lot more 
region gets illuminated due to such reflections when compared with specular 
reflection. Multiple diffuse reflections would 
lead to even larger spreads and it was not intuitively apparent as to how 
holes could be created in such widely spread visible regions. In fact, 
it seems unlikely that too many 
such holes can be created; we suspect that the number of holes in $V_2(s)$ is
$O(n)$.

\section{The construction of a hole for two diffuse reflections}
\label{constructhole}

\begin{figure}[htbp]
\centerline{\psfig{figure=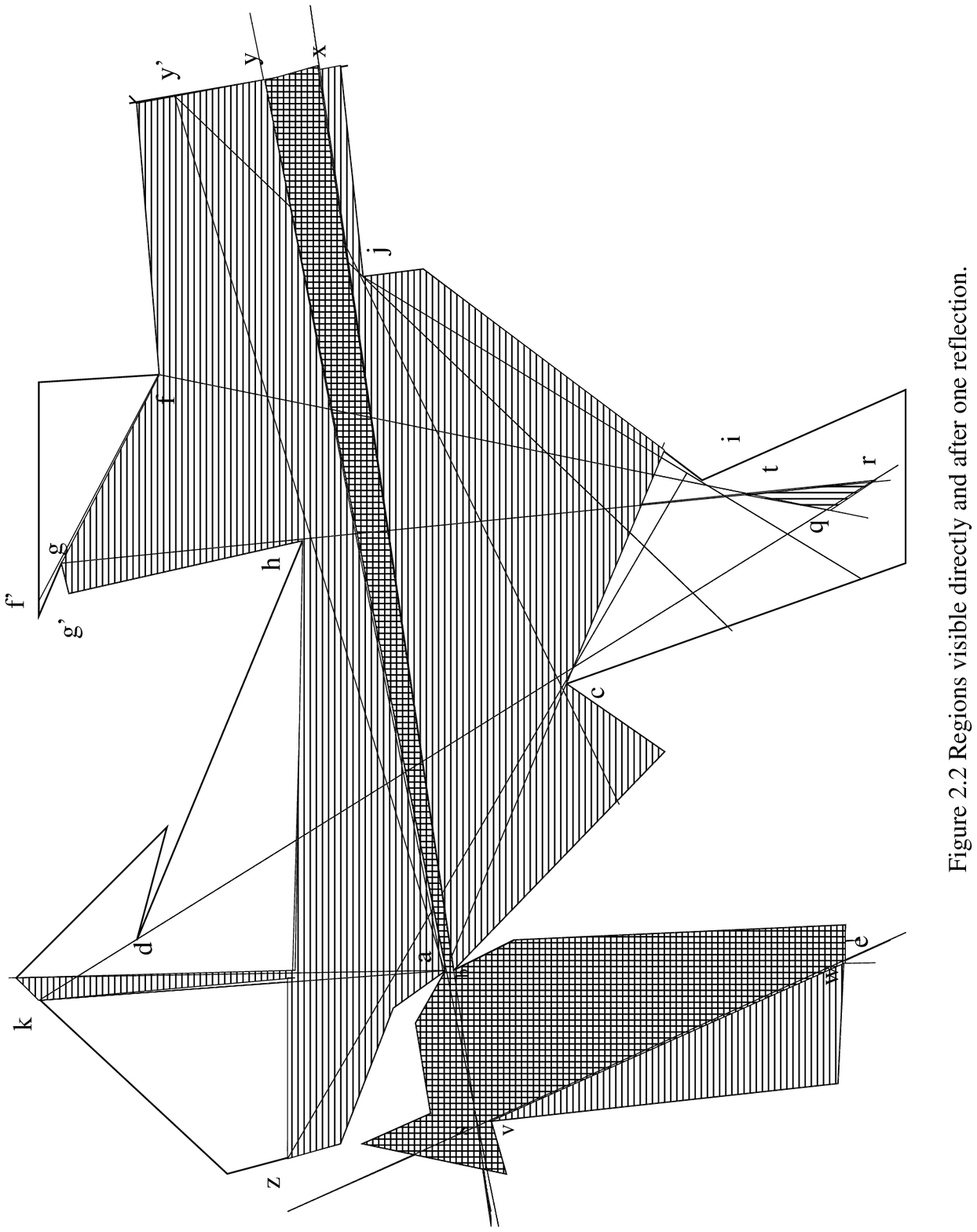}}
\caption{Existence of hole $tqr$ in $V_2(s)$}  
\label{figex3}
\end{figure}

\begin{figure}[htbp]
\centerline{\psfig{figure=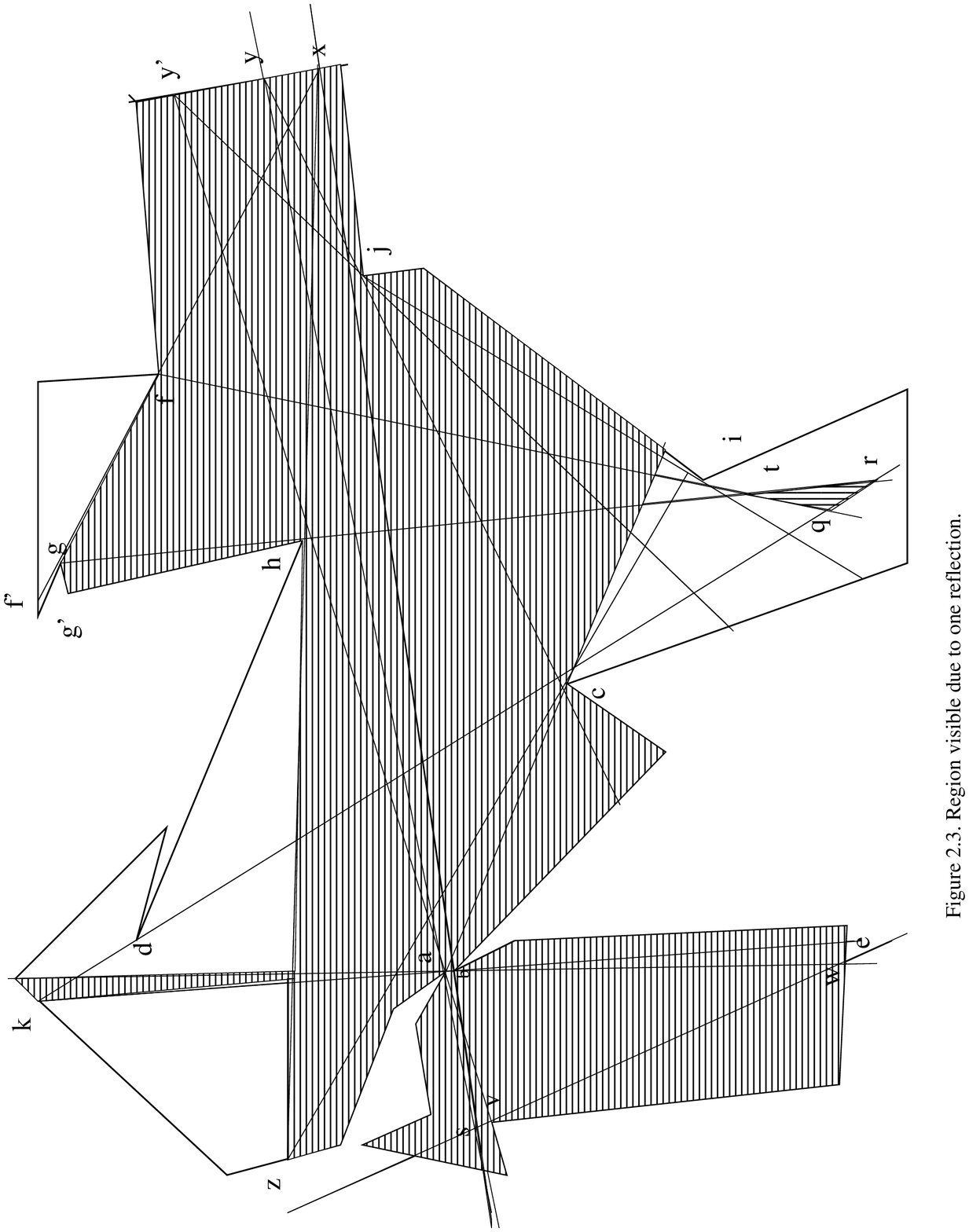}}
\caption{Existence of hole $tqr$ in $V_2(s)$}  
\label{figex2}
\end{figure}

Now we show that $V_2(s)$ may have a hole. 
In what follows, we state this result in detail as in \cite{sppdstrmay2001}.
We show the existence of an island or hole 
by constructing a simple polygon $P$ 
containing a point $s$ such that $V_2(s)$ has a triangular hole with
vertices $t$, $q$ and $r$ (see Figures \ref{figex3} and \ref{figex2}). 
Note that rays emanating from $s$ are drawn with dashed lines.
Rays that emerge after the first reflection (from points visible to $s$) 
are drawn with normal lines. Rays that emerge from points 
of $bd(V_1(s))$ (after the second reflection) are drawn with lines 
having alternating dots and dashes.
The interior of the triangle $tqr$ is invisible from $s$,
from all points of $bd(V_0(s))$, and, from all points on $bd(V_1(s))$.
Note that the first reflection takes place on points of 
$bd(V_0(s))$ and the second reflection takes place on points of 
$bd(V_1(s))$.

We need a definition. We say that a point $p$ {\it blocks} a 
region $R$ of $P$ from 
another region $S$ if the shortest path from every point 
in $R$ to every point in $S$ passes through $p$.


Clearly, no ray can reach the hole directly 
from $s$. This is because $b$ blocks $s$ from the (proposed) hole.  

\begin{lemma}
\label{1streflection}

The triangle $tqr$ is not visible from $s$ after the first reflection.

\end{lemma}

\begin{proof}

Consider $bd(V_0(s))$. 
We start by classifying portions of $bd(P)$ that form  
$bd(V_0(s))$. There are three portions- $bd(x,y)$, $bd(w,b)$ and $bd(a,v)$.
These constitute all points on $bd(V_0(s))$. 
The sequences of points,  $<s,a,y>$, $<s,b,x>$ and $<s,v,w>$ are collinear. 
How do we guard the hole from the points of $bd(V_0(s))$? 
The shortest path 
from any point in $bd(x,y)$ to any point in the hole has to pass through $j$.
In other words, $j$ $blocks$ the hole from $bd(x,y)$. 
So, no point in $bd(x,y)$ can see the hole. 
Similarly, $b$ blocks $bd(w,b)$ from the holes, and, 
$c$ blocks $bd(a,v)$ from the hole. So, no point on 
$bd(V_0(s))$, can see the hole.
We conclude that the hole is not visible from any point of $bd(V_0(s))$.

\end{proof}

All we need to do now is to show that $bd(V_1(s))$ can not see the hole.
We start by classifying portions of $bd(P)$ that form  $bd(V_1(s))$.
These are

\begin{enumerate}

\item Points of $bd(a,b)$ visible from $bd(a,v)$, $bd(w,b)$
or $bd(x,y)$. (Note that $bd(x,y)$ is the
only portion of $bd(P)$ visible from $s$ outside $bd(a,b)$.)

\item Points of $bd(i,f)$ and $bd(b,c)$, visible from $bd(a,v)$. (Points of $bd(a,v)$ 
can see points within $bd(i,f)$ because (i) the vertex $i$ is to the
right of the ray starting at $a$ and passing through $c$, and, 
(ii) the vertex $f$ is to 
the left of the ray starting at $v$ and passing through $a$. Observe that 
$bd(a,v)$ can also see points on $bd(b,c)$. These are the 
only portions of $bd(P)$, visible from $bd(a,v)$. 
)

\item Points of $bd(h,a)$ visible from $bd(w,b)$. 
(Points in $bd(w,b)$ can see points of
$bd(P)$ only within $bd(d,a)$, or more precisely, points within 
$bd(d,k)$, where  $e,b,a$ and $k$ are collinear.)

\item Points of $bd(b,c)$ visible from $bd(x,y)$. (The point $c$ is to the
left of $ray(y,j)$. So, all points of $bd(V_1(s))$ on $bd(b,j)$ are
to the right of $ray(y,j)$).

\item Points of $bd(h,a)$ visible from $bd(x,y)$.

\item Points of $bd(i,h)$ visible from $bd(x,y)$.

\end{enumerate}

The remaining points on $bd(P)$ can not be on $bd(V_1(s))$.

Now we show that none of these six possible portions of 
$bd(V_1(s))$ can illuminate the 
triangular hole $tqr$ of $V_2(s)$. 

\begin{enumerate}

\item The second reflection at $bd(a,b)$ causes rays to 
miss the hole due to the blocker $c$. This is due to the fact that
the shortest path from any point in $bd(a,b)$ to any point in the hole 
$tqr$, has to cross segment $ac$. 
In particular, points in $bd(v,b)$ are blocked from the hole by $b$.

\item Consider points of $bd(V_1(s))$ on $bd(i,f)$. It 
is sufficient to consider the extreme points 
$y'$ and $c'$ of $bd(i,f)$ in $bd(V_1(s))$. The point $y'$  
is collinear to $v$ and $a$. 
The point $j$ blocks $y'$ and the whole of $bd(j,y')$ 
from the hole. The point $c'$ is collinear with $a$ and $c$.
The point $i$ blocks the whole of $bd(i,j)$ from the hole $tqr$.

Now consider points of $bd(b,c)$ that belong to $bd(V_1(s))$ due to first reflection
at $bd(a,v)$. These points are blocked from the hole by $c$.

\item Only points within $bd(d,k)$ in $bd(h,a)$ can cause the second 
reflection from rays undergoing the first reflection on 
$bd(w,b)$. All these points in $bd(d,k)$ are blocked by $d$ i
from the hole 
$tqr$. Indeed, $k$, $d$, $q$ and $r$ are collinear, defining the edge 
$qr$ of $tqr$. Note that $q$ is the intersection of $ray(k,d)$ and
$ray(f,i)$ and, $r$ is the intersection of $ray(k,d)$ and $ray(g,h)$. 

\item Consider points of $bd(b,c)$ visible from $xy$. 
Points of $bd(b,c)$ 
can not see $tqr$ because they are blocked by $c$. 

\item Consider $bd(d,a)$. Points in $bd(d,z)$ are blocked by 
$h$ from points on $xy$. Since $z$, $h$ and $x$ are collinear and $tqr$ is
to the right of $ray(z,c)$, all points of $bd(z,a)$ are blocked from $tqr$
by $c$. 

\item Consider $bd(i,h)$ and $xy$. Points on $bd(j,y)$ are blocked from
$tqr$ by $j$.
So we
need to consider only $bd(y,h)$. Any point on $bd(y,f)$ may 
reflect into the region to the right of $ray(f,i)$ 
but not to the left of
$ray(f,i)$. In the construction,  
$f$, $i$, $t$ and $q$ are collinear and $tq$ defines an
edge of the hole $tqr$. Note that $ray(g,h)$ and $ray(f,i)$ intersect at
vertex $t$. 

Now consider point $f'$ on $bd(f,h)$ where  $f'$ is the point 
where $ray(x,f)$ meets $bd(P)$.
We construct $bd(f,g)$ so that the shortest path from $f'$ to $h$ passes 
through $g$ and $g$ lies in the segment joining $f$ and $f'$.
Points on $xy$ can not see any point in the interior of $bd(f,f')$. Since
$g$, $h$, $t$ and $r$ are collinear, (i) the shortest path 
from any point 
in $bd(f',g)$ to any point inside $tqr$ has to pass 
through blockers $h$ and $g$, and, (ii) the shortest path from any point in
$bd(g,h)$ to any point in $tqr$  has to pass through blocker $h$.
Indeed, $ray(g,h)$ defines the edge $tr$ of hole $tqr$.

Note that reflections on edge $g'h$ will not reach the interior of
triangle $tqr$ because triangle $tqr$ is to the right of
$ray(g',h)$.

\end{enumerate}

In summary, we have established that $tqr$ is a hole in $V_2(s)$. We state the 
result as a theorem.

\begin{theorem}
\label{2ndreflection}
It is possible to construct a simple polygon $P$ with  
internal points $s$, $t$, $q$ and $r$ such that (i) $tqr$ is a triangle
totally contained in $int(P)$ 
(ii) all points inside the triangle $tqr$ belong to $P\setminus V_2(s)$ and 
(iii) $tq$, $tr$  and $rq$ are on $bd(V_2(s))$.

\end{theorem}

\section{Concluding remarks}

We have shown that the region visible due to at most 
two diffuse reflections can have holes. This result holds for 
single 
point light sources as well as single extended light sources like a 
line segment or a circle. 
The details are simple and are omitted here. 
For three or more diffuse reflections too, such multiply connected
visible regions can be created by extending our construction; the
source $s$ in our construction can be created as an illuminating
line segment on an edge of the
polygon after a predermined number $k-2, k\geq 3$ of
reflections from a point $s'$, where light from $s'$ must undergo
$k-2$ reflections (inside say, a winding spiral) 
and hit $s$ before spreading to the rest of the polygon. 
The details 
are straighforward and are omitted here. 

An important research direction is that of establishing 
upper bounds on the number of holes in diffuse visibility
regions. We believe that the upper bound is as low as linear 
in the number of vertices of the input polygon.

\bibliography{newbib.bib}

\begin{thebibliography}{10}

\bibitem{addpp982}
B.~Aronov, A.~Davis, T.~K. Dey, S.~P. Pal, and D.~C. Prasad.
\newblock Visibility with multiple reflections.
\newblock {\em Discrete and Computational Geometry}, 20:61--78, 1998.

\bibitem{addpp981}
B.~Aronov, A.~Davis, T.~K. Dey, S.~P. Pal, and D.~C. Prasad.
\newblock Visibility with one reflection.
\newblock {\em Discrete and Computational Geometry}, 19:553--574, 1998.

\bibitem{g91}
S.~K. Ghosh.
\newblock Computing the visibility polygon from a convex set and related
  problems.
\newblock {\em Journal of Algorithms}, 12:75--95, 1991.

\bibitem{ea81}
H.~El Gindy and D.~Avis.
\newblock A linear algorithm for computing the visibility polygon from a point.
\newblock {\em Journal of Algorithms}, 2:186--197, 1981.

\bibitem{ghlst87}
L.~Guibas, {J}. Hershberger, {D}. Leven, {M} Sharir, and {R}. Tarjan.
\newblock Linear time algorithms for visibility and shortest path problems
  inside triangulated simple polygons.
\newblock {\em Algorithmica}, 2:209--233, 1987.

\bibitem{l83}
D.~T. Lee.
\newblock Visibility of a simple polygon.
\newblock {\em Computer Vision, Graphics and Image Processing}, 22:207--221,
  1983.

\bibitem{ll86b}
D.~T. Lee and A.~K. Lin.
\newblock Computing the visibility polygon from an edge.
\newblock {\em Computer Vision, Graphics and Image Processing}, 34:1--19, 1986.

\bibitem{o87}
J.~O'Rourke.
\newblock {\em Art gallery theorems and algorithms}.
\newblock Oxford University Press, 1987.

\bibitem{sppdstrmay2001}
S.~P. Pal and D.~Sarkar.
\newblock On multiple-connectedness of regions visible due to multiple diffuse
  reflections.
\newblock {\em Technical Report No. TR/IIT/CSE/SPP1, May 2001, Department of
  Computer Science and Engineering, Indian Institute of Technology, Kharagpur,
  721302, India.}, May 2001.

\bibitem{ppd98}
D.~C. Prasad, S.~P. Pal, and T.~K. Dey.
\newblock Visibility with multiple diffuse reflections.
\newblock {\em Computational Geometry: Theory and Applications}, 10:187--196,
  1998.

\bibitem{s92}
T.~Shermer.
\newblock Recent results in art galleries.
\newblock {\em Proceeding of the IEEE}, 80:1384--1399, September 1992.

\end{thebibliography}
\end{document}